\documentclass[twocolumn,aps,longbibliography, pra,showpacs,groupedaddress,superscriptaddress,nofootinbib]{revtex4-1}

\usepackage[utf8]{inputenc}  
\usepackage[T1]{fontenc}

\usepackage[normalem]{ulem}

\usepackage[dvipsnames]{xcolor}
\usepackage{graphicx,epsfig}
\usepackage{braket}
\usepackage{color}
\usepackage{amsmath}
\usepackage{amsfonts}
\usepackage{fancyhdr}
\usepackage{mathrsfs}
\usepackage[makeroom]{cancel}
\usepackage{mathtools}

\usepackage{subfigure}
\usepackage{float}

\usepackage[pdftex,linkcolor=blue,citecolor=blue,urlcolor=blue,colorlinks]{hyperref}

\renewcommand{\braket}[2]{\langle#1\vert#2\rangle}
\newcommand{\mean}[1]{\langle#1\rangle}

\DeclarePairedDelimiter\abs{\lvert}{\rvert}%
\DeclarePairedDelimiter\norm{\lVert}{\rVert}%
\makeatletter
\let\oldabs\abs
\def\abs{\@ifstar{\oldabs}{\oldabs*}}
\let\oldnorm\norm
\def\norm{\@ifstar{\oldnorm}{\oldnorm*}}
\makeatother

\renewcommand{\vec}[1]{\mbox{\boldmath$#1$}}

\newcommand{\enum}[1]{\mathit{e}^{#1}}
\newcommand{\hamdens}{\hat{\mathcal{H}}}
\newcommand{\ham}{\hat{H}}
\newcommand{\Spvek}[2][r]{%
  \gdef\@VORNE{1}
  \left(\hskip-\arraycolsep%
    \begin{array}{#1}\vekSp@lten{#2}\end{array}%
  \hskip-\arraycolsep\right)}
\newcommand{\appropto}{\mathrel{\vcenter{
\offinterlineskip\halign{\hfil$##$\cr
  \propto\cr\noalign{\kern2pt}\sim\cr\noalign{\kern-2pt}}}}}

\begin{document}
\title{Coherent spin mixing via spin-orbit coupling in Bose gases}
\author{J. Cabedo}
\affiliation{Departament de F\'isica, Universitat Aut\`onoma de Barcelona, E-08193 Bellaterra, Spain.}
\author{J. Claramunt}
\affiliation{Departament de Matem\`atiques, Universitat Aut\`onoma de Barcelona, E-08193 Bellaterra, Spain.}
\author{A. Celi}
\affiliation{Departament de F\'isica, Universitat Aut\`onoma de Barcelona, E-08193 Bellaterra, Spain.}
\author{Y. Zhang}
\affiliation{International Center of Quantum Artificial Intelligence for Science and Technology and Department of Physics, Shanghai University, Shanghai 200444, China}
\author{V. Ahufinger}
\affiliation{Departament de F\'isica, Universitat Aut\`onoma de Barcelona, E-08193 Bellaterra, Spain.}
\author{J. Mompart}
\affiliation{Departament de F\'isica, Universitat Aut\`onoma de Barcelona, E-08193 Bellaterra, Spain.}

\begin{abstract}
We study beyond-mean-field properties of interacting spin-1 Bose gases with synthetic Rashba-Dresselhaus spin-orbit coupling at low energies. We derive a many-body Hamiltonian following a tight-binding approximation in quasi-momentum space, where the effective spin dependence of the collisions that emerges from spin-orbit coupling leads to dominant correlated tunneling processes that couple the different bound states. We discuss the properties of the spectrum of the derived Hamiltonian and its experimental signatures. In a certain region of the parameter space, the system becomes integrable, and its dynamics becomes analogous to that of a spin-1 condensate with spin-dependent collisions. Remarkably, we find that such dynamics can be observed in existing experimental setups through quench experiments that are robust against magnetic fluctuations.
\end{abstract}
\maketitle


\section{Introduction}

Over the last decade, synthetic gauge fields have been experimentally realized in neutral atom systems \cite{Dalibard-2011, Goldman-2014}, which provide a highly controllable and tunable platform for quantum many-body simulations \cite{Maciej-Book-2012}. The achievement of Bose-Einstein condensate (BEC) with spin-orbit coupling (SOC) by the NIST group \cite{Lin-2011} gave rise to a huge body of theoretical and experimental research, particularly focusing on spin-1/2 systems. Such spin-orbit coupled gases are characterized by parameter-dependent nontrivial single-particle dispersion relations. In an interplay with the inter-atomic interactions, these yield a rich phase diagram, including a zero-momentum phase, a spin-polarized phase and a spatially modulated phase with supersolid-like properties \cite{Li-2012, Li-2013, Li-2017,Bersano-2019} (for supersolid phases from magnetic interactions see \cite{Tanzi-2019, Bottcher-2019, Chomaz-2019}). Likewise, the presence of SOC notably affects the dynamics of the gas, with an excitation spectrum exhibiting peculiar features such as anisotropy, suppression of the sound velocity or the emergence of a roton minimum in the plane-wave phase \cite{Zhang-2012, Khamehchi-2014, Ji-2015, Khamehchi-2017}. 

\begin{figure*}[t!]
\includegraphics[width=0.95\linewidth]{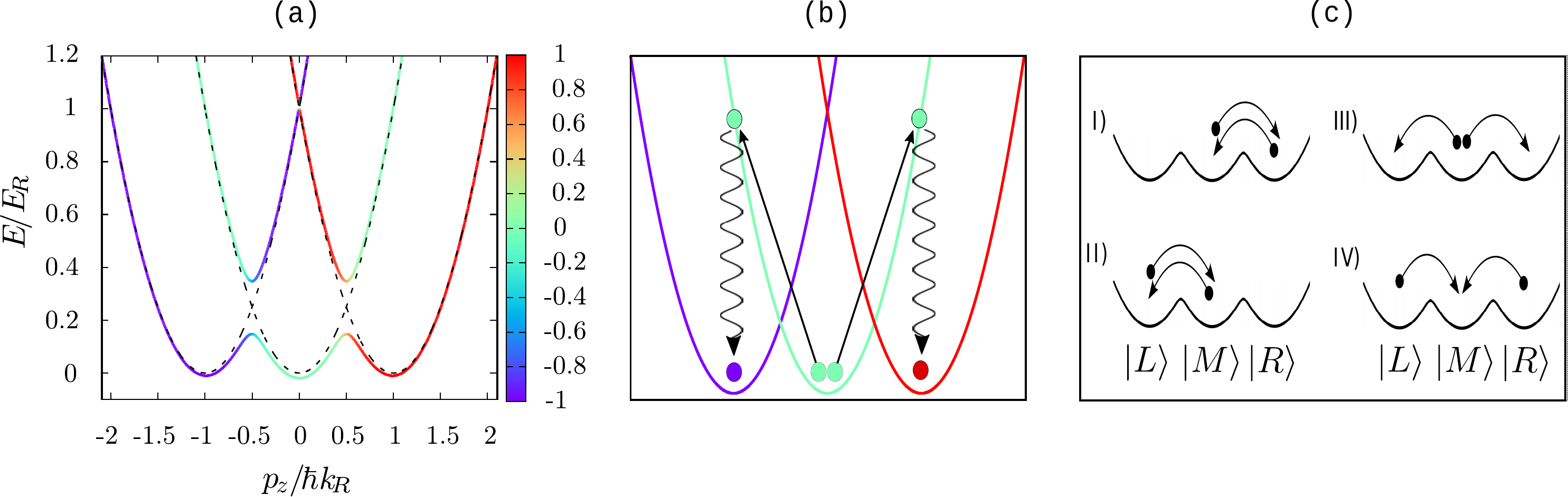}
\caption{\small (Color online) \textbf{Emergence of spin-changing collisions in presence of SOC.}  \textbf{(a)} Dispersion bands of Hamiltonian \eqref{hamspnotrap2} along the longitudinal direction $\hat{z}$ for $\Delta_s = 0$, and $\hbar\Omega=0.2 E_R$. The color texture represents the expected value of the spin of the  dressed states. Dashed lines show the undressed dispersion bands (at $\Omega = 0$). \textbf{(b)} Schematic representation of a resonant collision process that couples different spin states mediated by Raman transitions (represented in wavy lines). \textbf{(c)} Spin-changing collision processes that couple the many-body states in the Fock space spanned by the tight-binding well-states basis $\ket{L}$, $\ket{M}$ and $\ket{R}$ (see section \ref{section3}), which act as effective correlated tunneling processes between the bound states.}\label{spinchangingcollisions} 
\end{figure*}

In the simplest scenario, the engineered SOC consists of equal Rashba \cite{Bychkov-1984} and Dresselhaus \cite{Dresselhaus-1965} contributions. This restricted one-dimensional (1D) kind of SOC can be achieved by employing an external magnetic field and pairs of counter-propagating laser beams that couple different atomic states. These gases exhibit a rich phase diagram that results from the interplay of the number of minima of the dispersion bands and the nature of the interactions in the gas \cite{Zhang-2016}. More recently, Rashba SOC in two-dimensional (2D) BECs \cite{Wu-2016, Sun-2016} and in ultracold Fermi gases \cite{Huang-2016} was achieved.

While most of the research works focus on spin $1/2$ gases, spin-orbit coupled BECs with spin larger than 1/2 have been theoretically studied \cite{Lan-2014,Martone-2016,Sun-2016} and spin-1 BECs with SOC were attained \cite{Campbell-2016}. At the same time, the majority of the related research involves very dilute gases, where the interactions are weak and a mean-field treatment is accurate.
Yet many intriguing phenomena appear in presence of SOC beyond the mean-field regime. This is the case, for instance, in optical lattices where, by downplaying the kinetic terms and enhancing the gas density without further losses, many-body physics at strong coupling may become experimentally accessible \cite{Bloch-review-2008}. 
The experimental observation of integer quantum Hall (Hofstadter model on 2D square lattice \cite{Aidelsburger-2013,Miyake-2013}, and on narrow strips \cite{Atala-14,Stuhl-15,Mancini-15,Livi-16,Kolkowitz-17} in real or synthetic lattices \cite{Boada-12,Celi-2014,Ozawa-19}) and spin-Hall  (Haldane model in honeycomb-like lattices \cite{Jotzu-14,Flaschner-16}) effect for noninteracting gases with synthetic gauge fields, the lattice equivalent of SOC, paves the way to the experimental realization of fractional quantum Hall effect and quantum magnetism with interacting gases.
Remarkably, beyond-mean-field effects can dominate the dynamics in weakly-interacting dilute gases when mean-field effects largely cancel and lead to the stabilization of quantum droplets \cite{Petrov-15}, as experimentally demonstrated for contact \cite{Cabrera-2017,Cheiney-18,Semeghini-2018} and dipolar interactions \cite{Kadau-16,Schmitt-16,Chomaz-16}.

In this work, we show that beyond-mean-field effects can dominate the dynamics of weakly-interacting spin-$1$ Bose gases with Raman-induced artificial Rashba-Dresselhaus SOC. At weak couplings, the single-particle dispersion relation exhibits a triple-well shaped lowest band. Similarly as done in \cite{Higbie-2004} for the spin-$1/2$ gas, we consider the well-shaped band to act as a 3-site lattice in momentum space by performing a tight-binding approximation. Even in the case of $SU(3)$-symmetric interactions, we show that, for spin $1$ and larger, the SOC-mediated spin dependence of the interactions gives rise to the appearance of correlated tunneling processes involving the tightly-bound lowest-band states, as illustrated in Fig.\,\ref{spinchangingcollisions}, leading to a richer scenario than in the spin-$1/2$ case.

Synthetic momentum space lattices can be obtained also via Bragg transitions in a BEC, as proposed in \cite{Gadway-2015} and experimentally realized  \cite{Meier-2016,An-17,Meier-18} to simulate topological models. As shown in \cite{Fangzhao-2018},  Bose statistics can induce localized interactions in momentum space at the mean-field level. As we show here, nontrivial many-body physics in SOC-induced momentum space lattices emerges due to the interplay of contact binary collisions with the spin texture present in the dispersion band of a SOC Bose gas. 
The work is structured as follows. In section \ref{section2}, we review the Hamiltonian for an atomic gas with SOC and introduce a second-quantized form for the weakly-interacting bosonic gas in a lowest-band approximation. In section \ref{section3}, we investigate the system in the weak coupling regime where the lowest single-particle dispersion band exhibits a triple-well shape. The spin dressing in the band gives rise to effective spin-changing collision processes. Within a tight-binding approximation in the lowest band, we show that such processes act as correlated tunneling terms between the bound states. In section \ref{section4}, we explore the properties of the momentum-space tight-binding Hamiltonian. For a particular set of parameters, the Hamiltonian becomes integrable. In this situation, the pseudo-spin dynamics is analogous to that of spin-1 BEC with spin-dependent collisions \cite{Law-1998}, where coherent spin-mixing is induced by nonlinear processes \cite{Zhang-2005,Chang-2005}. Finally, we discuss the feasibility of the model in state-of-the-art experiments.

\section{Physical system}\label{section2}

\subsection{Synthetic SOC}\label{section2.1}

We consider a dilute Bose gas subject to an external uniform bias magnetic field where three hyperfine states $\left\{\ket{F,m_F} \right\}$ of a given manifold $F\geq 1$ are coupled by two-photon Raman processes, as realized in \cite{Campbell-2016}. Here we label the targeted bare or uncoupled hyperfine states, which work as an effective spin basis, as $\ket{s}$, with $s\in\left\{-1,0,1\right\}$. Each two-photon process involves a momentum exchange between the Raman fields and the atoms along the $\hat{z}$ direction given by $\hbar k_R\vec{e}_z$. This defines the energy scale of the system through the two-photon recoil energy $E_{R} = \hbar^2 k^2_R/2m$\,\footnote{To avoid confusion, note that in the literature it is common to employ the single-photon recoil as the energy scale, instead of our choice made here, which would correspond to  $E_{R}/4$.}. The Raman dressing supposes translational symmetry breaking, by establishing a preferred frame. However, the single-particle Hamiltonian adopts a translationally invariant and time-independent form in a frame corotating and comoving with the laser fields \cite{Lan-2014}
\begin{align}\label{hamspnotrap2}
\hamdens_{0}(\vec{p}) &=  \frac{1}{2m}\left(p_z-\hbar k_R \hat{F}_z\right)^2 + \frac{\vec{p}_{\perp}^2}{2m} + \frac{\hbar \Omega}{2}\hat{F}_x \nonumber \\
&\quad + \sum_s \hbar\Delta_s\ket{s}\bra{s},
\end{align}
where $\vec{p}_{\perp} = p_x\vec{e}_x + p_y\vec{e}_y$ and $(\hat{F}_x,\hat{F}_y,\hat{F}_z)$ are the spin-$1$ Pauli matrices. In Hamiltonian \eqref{hamspnotrap2}, the rotating wave approximation is considered, with $\Omega$ being the two-photon Rabi frequency of the Raman processes. The spin states energy shifts, $\hbar\Delta_s$, can be independently adjusted by controlling the detunings of the Raman lasers \cite{Campbell-2016}. Hamiltonian \eqref{hamspnotrap2} effectively describes a free spin-1 Bose gas with Rashba-Dresselhaus SOC. However, notice that in this case the SOC canonical and mechanical momentum differ by  $s \hbar k_R$.

The SOC term appearing in \eqref{hamspnotrap2} is given by
\begin{equation}\label{RDSO}
\hamdens_{RD} = -\gamma p_z\hat{F}_z,
\end{equation}
with a SOC strength $\gamma = \frac{\hbar k_R}{m}$. This term gives a linear contribution in $p_z$ to the dispersion relation, in a way that depends on the effective spin of the particle. By construction, the 1D SOC Hamiltonian \eqref{RDSO} breaks parity symmetry. Instead, it possesses invariance under the simultaneous action of parity and spin-flip operation, which we will refer to as s-parity symmetry. It is worth mentioning that this is not generally the case in artificial SOC, which can be tailored in a way where each spin component is coupled independently to the momentum degree of freedom. For simplicity, we have restricted Hamiltonian \eqref{hamspnotrap2} to the scenario in which Raman momentum transfers and Rabi frequencies are set equal. In this setting, s-parity symmetry is maintained in the whole system by further fixing $\Delta_{-1} = \Delta_{1}$.

\subsection{Momentum-space triple-well band}\label{section2.2}

The spectrum of \eqref{hamspnotrap2} is characterized by three dispersion bands, denoted by $h_0(p_z)$, $h_1(p_z)$ and $h_2(p_z)$, along the direction of the Raman momentum transfers. For a nearly resonant SOC in the weak-coupling regime, $\hbar\Omega, \hbar\abs{\Delta_s} \ll E_R$, the bands for the different spins hybridize only around the crossings, which are turned into avoided crossings with a gap increasing with $\Omega$. This results into a triple-well shape of the lowest band, as illustrated in Fig.\,\ref{spinchangingcollisions}(a).

We consider the spin-orbit coupled gas to be spatially confined by means of an internal-state-independent potential $\hat{V}_t = \hat{V}_z(z) + \hat{V}_{\perp}(\vec{r}_{\perp})$, with $\vec{r}_{\perp} = x\vec{e}_x + y\vec{e}_y$. We will consider the longitudinal potential to be quadratic in $z$, that is $\hat{V}_{z} =  \frac{1}{2}m\omega_z^2\hat{z}^2$. In momentum space, the harmonic potential acts as an effective kinetic-like term in the single-particle Hamiltonian, being proportional to the second derivative of the momentum, which prevents the solutions of the system from being well-localized. It will be useful to write the Hamiltonian of the trapped single-particle system, $\hamdens_{s.p.} = \hamdens_0 + \hat{V}_t$, in the eigenbasis of the homogeneous Hamiltonian \eqref{hamspnotrap2}, the so-called dressed basis. Labelling the dressed states as $\{\ket{\varphi_0 (\vec{p})},\ket{\varphi_1 (\vec{p})}, \ket{\varphi_2 (\vec{p})}\}$, we can write
\begin{align}\label{fullhamsp}
\hamdens_{s.p.}(\vec{p}) = &\sum_i \left( h_i(p_z)+ \frac{\vec{p}^2_{\perp}}{2m}\right) \ket{\varphi_i(\vec{p})}\bra{\varphi_i (\vec{p})} \nonumber \\
& -\frac{1}{2}m\hbar^2\omega_z^2 \hat{\text{U}}^{\dagger}(p_z) \frac{\partial^2}{\partial p_z^2}\hat{\text{U}}(p_z) + \hat{V}_{\perp},
\end{align}
where $\hat{\text{U}}(p_z) = \sum_{s,j} U_{s,j}(p_z) \ket{s,\vec{p}} \bra{\varphi_j(\vec{p})}$ is the unitary transformation that relates the dressed basis with the uncoupled hyperfine state basis $\left\{\ket{s,\vec{p}}\right\}$, with $U_{s,j}(p_z) = \braket{s,\vec{p}}{\varphi_j(\vec{p})}$.

For our purposes, we require that the longitudinal trapping energy $\hbar \omega_z$ is significantly small compared to the energy split between the two lowest bands, that is, $\omega_z \ll \Omega$. In this weak longitudinal trapping regime, a lowest-band approximation can be safely applied: we truncate the single-particle basis to the lowest energy band states $\left\{ \ket{\varphi_0 (\vec{p})} \right\}$. Such basis states have at each quasi-momentum $\vec{p}$ an internal state composition \begin{equation}
\overrightarrow{s_0}(p_z) = \sum_{s} U_{s,0}(p_z)\ket{s}    
\end{equation} that depends on the strength of the Raman couplings and the detunings. For a state in the lowest band $\overrightarrow{\phi}(\vec{p}) = (\phi(\vec{p}),0,0)^T$, the energy due to the trapping in the $\hat{z}$ direction is given by
\begin{align}\label{trappingcontribution}
\mean{\hat{V}_z}_{\overrightarrow{\phi}} &= -\frac{1}{2}m \hbar^2 \omega_z^2 \overrightarrow{\phi}^{\dagger}\hat{\text{U}}^{\dagger}(p_z) \frac{\partial^2}{\partial p_z^2}\left(\hat{\text{U}}(p_z) \overrightarrow{\phi}\right) \nonumber \\
&= -\frac{1}{2}m \hbar^2 \omega_z^2\phi^*\left[\frac{\partial^2}{\partial p_z^2}-\norm{\frac{\partial \overrightarrow{s_0}}{\partial p_z}(p_z)}^2 \right]\phi.
\end{align}

From equations \eqref{fullhamsp} and \eqref{trappingcontribution} it follows that, in the lowest-band approximation, the single-particle Hamiltonian is reduced to $\hamdens_{s.p.} = \hamdens_z + \hamdens_{\perp}$, with
\begin{equation}
\hamdens_{\perp} =  \frac{\vec{p}^2_{\perp}}{2m} + \hat{V}_{\perp},
\end{equation}
and
\begin{equation}
\hamdens_z \simeq h(p_z) - \frac{1}{2}m\hbar^2\omega_z^2 \frac{\partial^2}{\partial p_z^2}.
\end{equation}
Here, $h = h_0 + \frac{1}{2}m \hbar^2 \omega_z^2 \norm{\frac{\partial \overrightarrow{s_0}}{\partial p_z}}^2$ is the effective energy band in the trapped system. When the confinement is weak compared to the recoil energy, the deviation from the free-particle band near the minima is negligible.

\subsection{Many-body Hamiltonian}\label{section2.3}
We now construct the many-body Hamiltonian by introducing the corresponding field operators for the band modes $\hat{\varphi}_j(\vec{p})$ obeying standard bosonic commutation relations: $\left[\hat{\varphi}_j(\vec{p}), \hat{\varphi}_k^{\dagger}(\vec{p}')\right] = \delta(\vec{p}-\vec{p}')\delta_{j,k} $. We write
\begin{equation}\label{manybodyham}
\ham = \ham_{n.i.} + \ham_{int},     
\end{equation}
where $\ham_{n.i.}$ and $\ham_{int}$ stand for its noninteracting and interacting contributions, respectively. In the lowest band approximation, the former is simply given by
\begin{align}\label{niham}
\ham_{n.i.} &\simeq \int d \vec{p} \hat{\varphi}_0^{\dagger}(\vec{p}) \hamdens_{s.p.} \hat{\varphi}_0(\vec{p}).
\end{align}

To derive $\ham_{int}$, we consider only binary s-wave contact collisions. For simplicity, we assume that they are $SU(3)$-symmetric and can, therefore, be characterized by a single parameter $g=\frac{4\pi\hbar^2a}{m}$. We take the state-independent scattering length $a$ to be positive. These considerations could apply, for instance, to the hyperfine states of $^{87}\text{Rb}$ \cite{Myatt-1996}. Nonetheless, a generalization to spin-dependent scattering lengths would be straightforward. 

In this way, expressed in the bare state basis, the interaction Hamiltonian is given by 
\begin{align}\label{interactpotential}
&\bra{s_1,\vec{p} ; s_2,\vec{p'}}\hamdens_{int}\ket{s_3,\vec{p''}; s_4,\vec{p'''}} \nonumber \\
&= \frac{g}{2(2\pi \hbar)^3}\delta_{s_1,s_4}\delta_{s_2,s_3}\delta(\vec{p}+\vec{p'}-\vec{p''}-\vec{p'''}),    
\end{align}
resulting in the following second-quantized form
\begin{widetext}
\begin{align}\label{inthamtotal}
\ham_{int} &= \frac{g}{2(2\pi \hbar)^3}\sum_{s_1,s_2} \int d\vec{p} d\vec{p'} d\vec{q} \hat{a}_{s_1}^{\dagger}(\vec{p}-\vec{q})\hat{a}_{s_2}^{\dagger}(\vec{p'}+\vec{q})\hat{a}_{s_2}(\vec{p'})\hat{a}_{s_1}(\vec{p}),
\end{align}
\end{widetext}
where $\hat{a}_{s}(\vec{p})^{\dagger}$, $\hat{a}_{s}(\vec{p})$ are the creation and annihilation operators for the mode $\ket{s,\vec{p}}$, respectively. As long as the energy per particle in the many-body system is lower than the band splitting, the lowest band approximation can be maintained. This is translated into an upper bound for the density in the gas
\begin{equation}\label{nmaxrecoil}
gn \ll E_R.
\end{equation}

In this situation, we can truncate the expression of $\hat{a}_{s}(\vec{p})$ to the lowest band
\begin{align}
\hat{a}_{s}(\vec{p}) = \sum_j U_{s,j}(p_z)\hat{ \varphi}_j(\vec{p}) \simeq U_{s,0}(p_z)\hat{\varphi}_0(\vec{p}).
\end{align}
After inserting this approximation into \eqref{inthamtotal} we obtain
\small \begin{align}\label{intham}
\ham_{int} &\simeq \frac{g}{2(2\pi \hbar)^3}\int d\vec{p} d\vec{p'} d\vec{q}\Bigg( \hat{\varphi}_0^{\dagger}(\vec{p}-\vec{q})\hat{\varphi}_0^{\dagger}(\vec{p'}+\vec{q})\nonumber\\
&\qquad\qquad \cdot\hat{\varphi}_0(\vec{p'})\hat{\varphi}_0(\vec{p})f(p_z,q_z)f(p_z',-q_z)  \Bigg),
\end{align}
\normalsize
with
\begin{align}\label{ffunction}
f(p_z,q_z) &= \sum_{s}U_{s,0}(p_z - q_z)U_{s,0}(p_z) \nonumber \\
&=\overrightarrow{s_0} (p_z-q_z)\cdot\overrightarrow{s_0} (p_z).
\end{align}

Remarkably, notice how despite assuming $SU(3)$-symmetric interactions, each scattering process $(\vec{p},\vec{p'}) \rightarrow (\vec{p}-\vec{q},\vec{p'}+\vec{q})$ in the lowest band is now weighted by the overlaps of the spin-states $\overrightarrow{s_0}$ of the initial and final states involved. Near the band minima, the spin overlaps decrease fast with its quasimomentum separation, which yields a certain degree of localization of the interactions in quasimomentum space. This feature is crucial, as it directly allows to drive correlated behavior in momentum space, while otherwise totally delocalized interactions could not \cite{Meier-2016}. Such phenomenon was exploited in \cite{Williams-2012} to create effective interactions with higher-order partial waves at low energies. Mediated by Raman photon pairs, particles can change spin via resonant collisions, as represented in Fig.\,\ref{spinchangingcollisions}(b), which leads to effective spin changing collisions. In the next section, we show that these processes can dominate the dynamics in a regime where a momentum space tight-binding approximation can be applied, in which only the lowest energy states of each well of the band are taken into account.

\section{Tight-binding approximation}\label{section3}

The degree of delocalization of the interactions in \eqref{intham} supposes a challenge to any immediately apparent truncation of the many-body Hilbert space. We now consider a tight-binding approximation in momentum space, in which we assume that the single-particle contributions to Hamiltonian \eqref{manybodyham} dominate. The interaction Hamiltonian acts then as a perturbation to the noninteracting system, and the low-energy scenario is well described within the Fock space spanned by the three lowest single-particle energy states, as long as the energy per particle is significantly smaller than the energy separation between such states and the next lowest family of energy eigenstates. 

In the weak longitudinal trapping regime, the wavefunctions of such states are localized in the vicinity of the minima of the wells in the band. The effective Hilbert space can be then truncated to just one single-particle state per site, the so-called well states,  with wave function $\psi_{i}(\vec{p})$. Here, $i\in \left\{-1,0,1\right\}$, which correspond to the left-, middle- and right-well states respectively. The transverse part of the wave function is spin independent, so it is useful to write $\psi_{i}(\vec{p})= \phi_{i}(p_z)\phi_{\perp}(\vec{p}_{\perp})$. The function $\phi_{i}(p_z)$ is centered around the corresponding minima at $p_z = p_i$, with $p_{\pm 1} \sim \pm \hbar k_R$ and $p_0 = 0$. Under these considerations 
\begin{equation}\label{truncated foperator}
\hat{\varphi}_0(\vec{p}) \sim \psi_{-1}(\vec{p})\hat{b}_{-1} + \psi_0(\vec{p})\hat{b}_0 + \psi_1(\vec{p})\hat{b}_1,
\end{equation}
where $\hat{b}_i$ is the bosonic annihilation operator for the $i^{\text{th}}$ well-state. At the vicinity of the minima, the dispersion along $\hat{z}$ is close to being quadratic and so one can treat each site as harmonic oscillators when $ \phi_{i}(p_z)$ is confined enough. This imposes a stricter upper limit on the density of the gas in the trap 
\begin{equation}\label{densityupbound2}
gn \ll \hbar\omega_z \ll E_R.
\end{equation}
In section \ref{section4.2} we discuss its experimental viability. Henceforth, we will assume that condition \eqref{densityupbound2} holds. With this simplification, the noninteracting contribution to the Hamiltonian in this low-energy description is reduced to (recall \eqref{niham})
\begin{equation}\label{fmniham}
\ham_{n.i.} \simeq \sum_{i}\epsilon_i\hat{N}_i -\frac{1}{2}\sum_{\left\langle i,j \right\rangle} J_{ij}\hat{b}_i^{\dagger}\hat{b}_j,
\end{equation}
with
\begin{align}
J_{ij} &= -2\int dp_z \phi_j^*(p_z) \hamdens_z \phi_i(p_z), \label{fmtunnelings}\\ 
\epsilon_i &= \int d\vec{p} \psi_i^*(\vec{p}) \hamdens_{s.p.} \psi_i(\vec{p}),\label{fmsiteenergies}\\
\Hat{N}_i &= \hat{b}_i^{\dagger}\hat{b}_i,
\end{align}
and where $\langle i,j \rangle$ stands for nearest neighbours summation.

Likewise, substituting the truncated field operator \eqref{truncated foperator} into \eqref{intham}, we obtain
\begin{equation}\label{inthamfmode}
\ham_{int} \simeq \sum_{i,j,k,l} U_{ijkl} \hat{b}_i^{\dagger}\hat{b}_j^{\dagger}\hat{b}_k\hat{b}_l,
\end{equation}
where the coefficients $U_{ijkl}$ are given by
\begin{equation}\label{intweights}
U_{ijkl} = \frac{g}{2(2\pi\hbar)}\int d\vec{r}_{\perp}\abs{\tilde{\phi}_{\perp}(\vec{r}_{\perp})}^4\int dq_z G_{il}(q_z)G_{jk}(-q_z),
\end{equation}

with
\small
\begin{equation}\label{gfunctions}
G_{ab}(q)= \int dp_z \phi_a^{*}(p_z-q_z)\phi_b(p_z)f(p_z,q_z).
\end{equation}
\normalsize
Here, $\tilde{\phi}_{\perp}$ is the inverse Fourier transform of the transverse mode $\phi_{\perp}$. It is easy to show that $G_{ab}(q) = G_{ba}(-q)$ for every index $a,b$, and hence
\begin{equation}\label{Usimmetries1}
U_{ijkl} = U_{jilk} = U_{klij} = U_{lkji}, \quad \text{for all }i,j,k,l.
\end{equation}
Furthermore from  s-parity symmetry it follows
\begin{align}\label{Usimmetries2}
U_{ijkl} &= U_{(-l)jk(-i)} = U_{i(-k)(-j)l} \nonumber \\
&= U_{(-j)(-i)(-l)(-k)}.
\end{align}

Condition \eqref{densityupbound2} implies that the wave functions $\phi_i(p_z)$ do not deviate significantly from the harmonic oscillator Gaussian states, with width $\sigma_z \simeq \sqrt{\frac{m\hbar\omega_z}{2}}$. From \eqref{intweights} and \eqref{gfunctions} it follows that, for $\sigma_z/(\hbar k_R) \ll 1$
\begin{equation}\label{Ucoefficients}
U_{ijkl} \approx C_{ijkl}(\Omega) U_0 \enum{-\Big( \frac{p_i + p_j - p_k - p_l}{4 \sigma_z}\Big)^2},
\end{equation}
with
\begin{equation}\label{ccoeff}
C_{ijkl}(\Omega) = \left(\overrightarrow{s_0} (p_i)\cdot\overrightarrow{s_0} (p_l)\right)\left(\overrightarrow{s_0} (p_j)\cdot\overrightarrow{s_0} (p_k)\right)  ,  
\end{equation}
and
\begin{equation}\label{U0coeff}
U_0 = \frac{g\mean{n}}{2 N}.
\end{equation}

Here, the coefficient $\mean{n}$ is the average density in the gas. Hence, as the longitudinal trapping frequency $\omega_z$ is made smaller, most coupling coefficients $U_{ijkl}$ decrease exponentially, while those relating modes $\{\phi_{i},\phi_{j}\}$ with $\{\phi_{i-k},\phi_{j+k}\}$, being $\abs{k} \in \{0,1,2\}$, decrease linearly. Hamiltonian \eqref{inthamfmode} can then be truncated to
\begin{align}\label{homfminthamshort}
\ham_{int} \simeq  \ham_{int}^{(0)} + &\sum_{i=0}^{1}\sum_{j=-1}^{0} 2U_{ij}^{(1)} \hat{b}_{i-1}^{\dagger}\hat{b}_{j+1}^{\dagger}\hat{b}_{i}\hat{b}_{j}\nonumber \\
+ &2U_{-1,1}^{(2)}\hat{b}_{1}^{\dagger}\hat{b}_{-1}^{\dagger}\hat{b}_{1}\hat{b}_{-1}
\end{align}
with
\begin{align}
\ham_{int}^{(0)} &:= \sum_{i,j} U_{ij}^{(0)}\hat{b}_i^{\dagger}\hat{b}_j^{\dagger}\hat{b}_i\hat{b}_j,\\
U_{ij}^{(k)} &:= U_{(i-k)(j+k)ji}. \label{Uijk}
\end{align}

The factors $C_{ijkl}(\Omega)$ \eqref{ccoeff} depend on the spin mixture of the band states around the minima. Specifically, on the spin projection of each pair of initial and final well states, which decreases fast with their inter-well distances $\abs{i-l}$ and $\abs{j-k}$. The gap opening between the two lowest dispersion bands depends linearly on $\Omega$, while the spin mixture around the minima increases quadratically. This allows to find a regime where the band gap is maintained sufficiently large compared to the energy scales of the system yet the spin overlap between distant sites is made arbitrarily small. Thus, the terms that correspond to zero momentum exchange, proportional to $U_{ij}^{(0)}$, are the largest contributions to the interaction Hamiltonian. However, as $\hbar\omega_z/E_R$ approaches $0$, the ratio $U_{ij}^{(0)}/U_{kl}^{(0)}$ approaches $1$, for all indices $i,j,k,l$. The corresponding terms then add up to a single one that depends only on the total number of particles, $N$
\begin{equation}\label{hamfmintzerothorder}
\ham_{int}^{(0)} \overset{\omega_z \rightarrow 0}{\longrightarrow} U_0 \hat{N}(\hat{N}-1). 
\end{equation}

Dropping this term and taking into account equalities $\eqref{Usimmetries1}$ and $\eqref{Usimmetries2}$, the total Hamiltonian in the tight-binding approximation can be approximated to
\begin{equation}\label{fmham1_v2}
\ham_{t.b.} = \ham_{n.i.} + \ham_{int} \simeq \ham_{1} + \ham_{2} + \hat{J},    
\end{equation}
with
\begin{align}
\ham_{1} &:= \,U_1\bigg[\hat{b}_L^{\dagger}\hat{b}_R^{\dagger}\hat{b}_M \hat{b}_M + \hat{b}_L\hat{b}_R\hat{b}_M^{\dagger}\hat{b}_M^{\dagger} \nonumber \\ &\qquad\quad +\hat{N}_R\hat{N}_M +\hat{N}_M\hat{N}_L \bigg] +  \epsilon\hat{N}_M  \label{Uoperator}\\
\ham_{2} &:= \,U_2\hat{N}_L\hat{N}_R, \label{U2operator}\\ 
\hat{J} &:= \,\frac{-J}{\sqrt{2}}\left(\hat{b}_R^{\dagger}\hat{b}_M + \hat{b}_L^{\dagger}\hat{b}_M + \hat{b}_R\hat{b}_M^{\dagger} + \hat{b}_L\hat{b}_M^{\dagger}\right). \label{Jperpoperator}
\end{align}
Here, to simplify the notation we have conveniently relabelled the left, middle and right well modes by identifiying $\left\{\hat{b}_{-1},\hat{b}_{0},\hat{b}_{1}\right\}$ with $\left\{\hat{b}_{L},\hat{b}_{M},\hat{b}_{R}\right\}$. Furthermore, due to s-parity symmetry we identify $U_1 := 2U_{RM}^{(1)} = 2U_{LM}^{(1)} = 2U_{MM}^{(1)} = 2U_{RL}^{(1)}$ (see \eqref{Uijk}), $J := J_{RM}/\sqrt{2} = J_{LM}/\sqrt{2}$ and $\epsilon_L = \epsilon_R$ (see \eqref{fmtunnelings} and \eqref{fmsiteenergies}). For convenience, the coefficient $2U_{RL}^{(2)}$ is relabelled as $U_2$. $\hat{J}$ can be interpreted as the trapping-mediated tunneling operator, while $\ham_{1}$ and  $\ham_{2}$ are the effective nearest neighbours and second-nearest neighbours interaction operators, respectively. Notice that due to the parity symmetry, the linear term in $N$ in \eqref{niham} reduces to an energy offset for the central well that we parametrize with $\epsilon = \epsilon_M - \epsilon_L$. We conveniently incorporate such term into $\ham_{1}$, despite being of noninteracting origin.

Thus, the operator $\ham_{1}$ naturally includes correlated tunneling terms proportional to $\hat{b}_L^{\dagger}\hat{b}_R^{\dagger}\hat{b}_M \hat{b}_M$ and $\hat{b}_L\hat{b}_R\hat{b}_M^{\dagger}\hat{b}_M^{\dagger}$ that couple the central well mode with the left and right modes. Remarkably, these are the leading order interaction terms in the tight-binding Hamiltonian \eqref{fmham1_v2}. Their presence clearly breaks the analogy of the quasimomentum space many-wells problem to a position space problem. Moreover, such processes involve more than two modes simultaneously, preventing its appearance in the most explored spin $1/2$ scenario \cite{Higbie-2004}. This crucial difference motivates the study here of the properties of the spin-1 system.

\section{Properties of the tight-binding Hamiltonian}\label{section4}

We now explore the properties of the tight-binding Hamiltonian \eqref{fmham1_v2} derived in the previous section, characterized by the presence of correlated tunneling terms that involve the bound states in momentum space.
\subsection{Spectral properties of \texorpdfstring{$\ham_{t.b.}$}{} }\label{section4.1}
\begin{figure*}[t]
\includegraphics[width=0.95\linewidth]{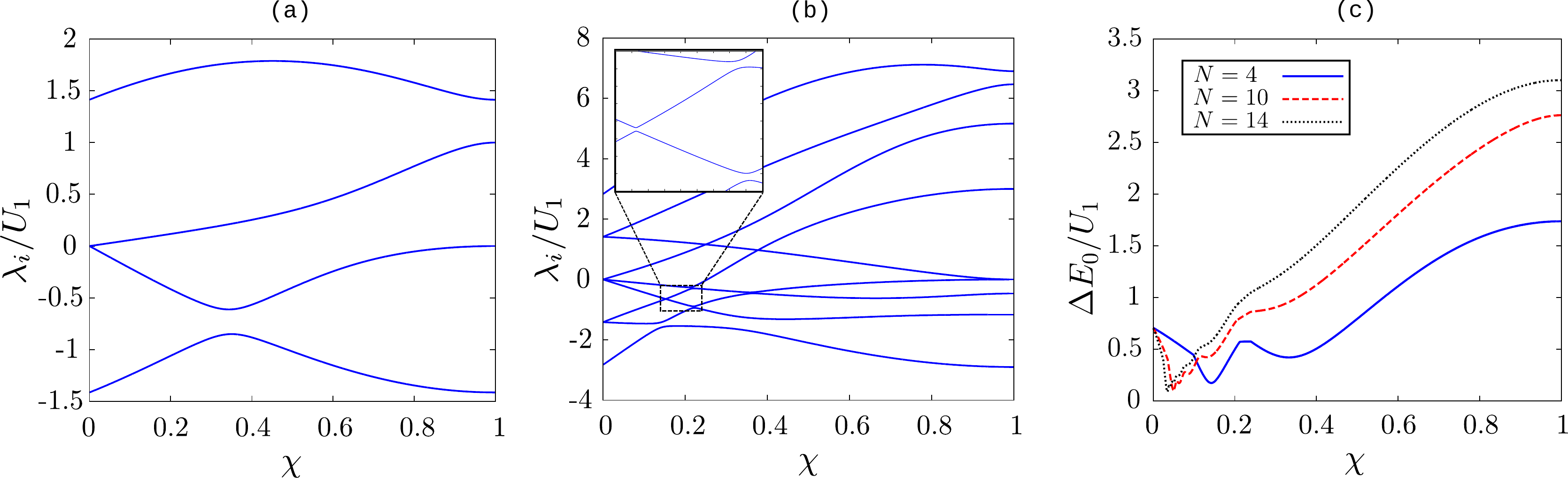}
\caption{\small (Color online) \textbf{Spectrum of the tight-binding Hamiltonian in the nonintegrable regime:} eigenvalues $\lambda_i$ of \eqref{fmham1_v2} with $\epsilon=0$ and  $U_2 = 0$, in the even parity subspace for \textbf{(a)} $N=2$ and \textbf{(b)} $N=4$, as a function of $\chi=\frac{2}{\pi}\arctan\left( \frac{U_1}{J} \right)$. The gaps at the avoided crossings are small but nonzero, as illustrated in the inset. The energy gap between the two lowest energy eigenstates is plotted in \textbf{(c)}, for $N=4$ (solid blue), $N=10$ (dashed red) and $N=14$ (dotted black).}\label{spectrumHS2HS4_eps0} 
\end{figure*}

We consider first the case $J/U_1 \ll 1$. Given the linear dependence on $\sigma_z$ that $U_1$ and $U_2$ acquire when $\omega_z \rightarrow 0$, a range of $\omega_z$ for which $J/U_1$ can be made arbitrarily small is guaranteed. The regime in which the interaction part of the Hamiltonian dominates is of interest, as all the properties of the Hamiltonian are related to its Raman-driven spin-orbit coupling nature. Moreover, we start by studying the situation were the second-nearest neighbour interactions can also be neglected, that is, when $U_2 \ll U_1$, which applies for small $\Omega$. Under these considerations
\begin{equation}\label{ham1storder}
\ham_{t.b.} \sim \ham_{1}, 
\end{equation}
which simply includes the possible collision processes between adjacent well states that exchange momentum, as shown in Fig.\,\ref{spinchangingcollisions}(c). Notice how such processes act as effective spin-changing collisions, as each well-state has a well-defined spin when the Raman coupling is weak. 

\begin{figure*}[t]
\centering
\includegraphics[width=0.95\linewidth]{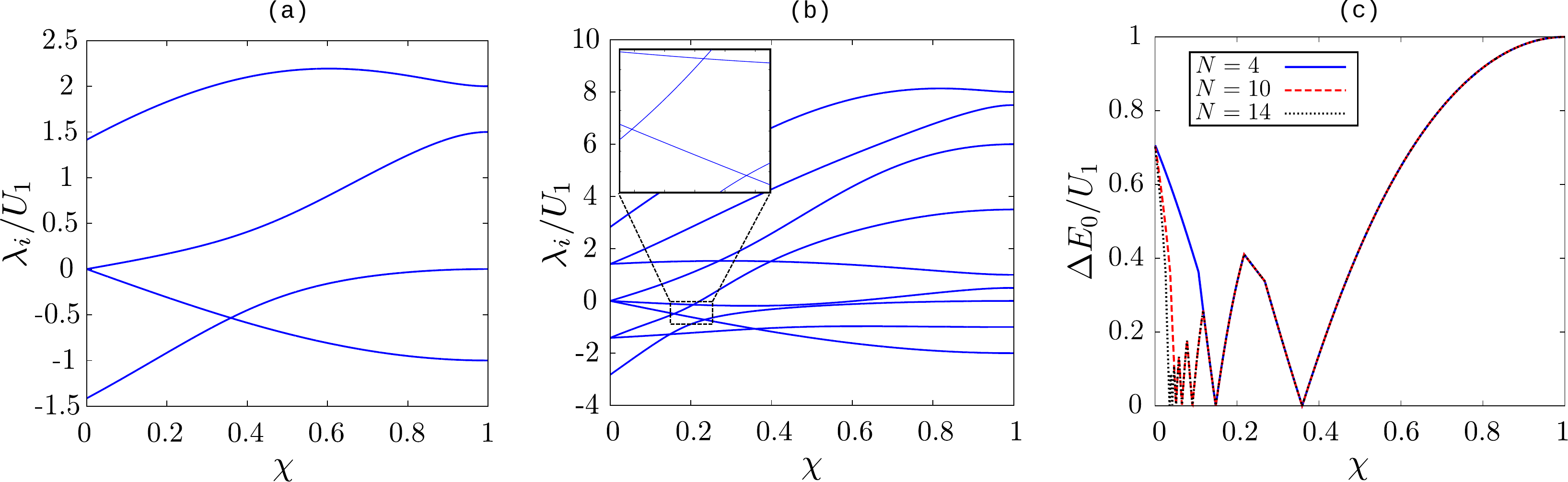}
\caption{\small (Color online) \textbf{Spectrum of the tight-binding Hamiltonian in the integrable regime:} eigenvalues $\lambda_i$ of \eqref{fmham1_v2} with $\epsilon=U_1/2$ and $U_2 = 0$, in the even parity subspace for \textbf{(a)} $N=2$ and \textbf{(b)} $N=4$, as a function of $\chi=\frac{2}{\pi}\arctan\left( \frac{U_1}{J} \right)$. The gaps at the crossings vanish at this value of the parameter $\epsilon$. The energy gap between the two lowest energy eigenstates is plotted in \textbf{(c)}, for $N=4$ (solid blue), $N=10$ (dashed red) and $N=14$ (dotted black).}\label{spectrumHS2HS4_eps05} 
\end{figure*}

Naturally, Hamiltonian $\ham_1$ possesses a $\text{U}(1)$ symmetry associated to the conservation of parity and total number of particles $N$. Moreover, the total spin is preserved in the collision processes, yielding an additional $\text{U}(1)$ symmetry associated to the conservation of the left- and right-well population imbalance, or magnetization. We conveniently define the magnetization operator as:
\begin{equation}\label{imbalanceop}
\hat{L}_z = \hat{N}_L - \hat{N}_R,
\end{equation}
with eigenvalues $m_z$. Furthermore, we show now that Hamiltonian $\ham_1$ acquires yet another $\text{U}(1)$ symmetry and becomes integrable when $\epsilon=U_1/2$. This is clear after the following transformation
\begin{align}\label{transfxy}
\hat{b}_x &= \frac{\hat{b}_L + \hat{b}_R}{\sqrt{2}}, \qquad \hat{b}_y = i\frac{\hat{b}_L - \hat{b}_R}{\sqrt{2}}.
\end{align}
With the operators $\hat{L}_x$ and $\hat{L}_y$ defined as
\begin{align}\label{JsJa}
\hat{L}_x& = \hat{b}_x^{\dagger}\hat{b}_M + \hat{b}_x\hat{b}_M^{\dagger}, \\
\hat{L}_y& = \hat{b}_y^{\dagger}\hat{b}_M + \hat{b}_y\hat{b}_M^{\dagger}, 
\end{align}
it is easy to check that $\left\{\hat{L}_x,\hat{L}_y,\hat{L}_z \right\}$ span the $\mathfrak{so}(3)$ Lie algebra, with $\left[\hat{L}_{\alpha} ,\hat{L}_{\beta}\right] = i\epsilon_{{\alpha}{\beta}{\gamma}}\hat{L}_{\gamma}$ for $\alpha,\beta,\gamma \in \left\{x,y,z\right\}$. Therefore, we can construct the Casimir element $\hat{L}^2 = \hat{L}^2_z + \hat{L}^2_y +\hat{L}^2_z$, and re-express Hamiltonian \eqref{Uoperator} as
\begin{align}\label{fmham1_v4}
\ham_{1} &= \frac{U_1}{2}\left[\hat{L}^2 -\hat{L}_z^2 + \left(\frac{2\epsilon}{U_1} -1\right)\hat{N}_M - \hat{N}\right].
\end{align}

Here, it is clear that the choice $\epsilon=U_1/2$ leaves the expression of the Hamiltonian only in terms of the operators $\hat{L}_\alpha$ and the total number of particles. The Hilbert space $\mathscr{H}$ can then be split into the orthogonal subspaces $\mathscr{H}^N_l$, corresponding to all the irreducible representations of $\mathfrak{so}(3)$ that are spanned in the subspace given by a total number of particles $N$, $\mathscr{H}^N$. These subspaces have dimension $2l+1$ and are labelled by the total number of particles $N$ and the quantum numbers $l$ associated to the $l(l+1)$ eigenvalues of $\hat{L}^2$. It is easy to prove that each subspace $\mathscr{H}^N$ realizes the $l=0,2,\dots,N$ representations when $N$ is even, and $l=1,3,\dots,N$ representations when $N$ is odd. In this most symmetric configuration, the spectrum of $\ham_{1}$ at $ \epsilon=U_1/2$ is fully characterized by the additional quantum number $m_z \in \left\{-l,-l+1,\dots,l\right\}$. The corresponding eigenvalues are given by
\begin{equation}\label{spectrumUeps05}
\lambda_{(N, l,m_z)} = \frac{U_1}{2}\left[l(l+1) - m_z^2 - N\right].    
\end{equation}

Notably, the $\mathfrak{so}(3)$ structure of $\ham_1$ at $\epsilon=U_1/2$ is preserved if we add a nonzero tunneling contribution to the Hamiltonian (recall \eqref{fmham1_v2}). Effectively, the tunneling operator expressed in the rotated basis \eqref{transfxy} reads $\hat{J} = -J\hat{L}_x$. However, in this case the $\text{U}(1)$ symmetry associated to the conservation of magnetization breaks down to a $\mathbb{Z}_2$ symmetry associated to parity conservation, and the integrability of the Hamiltonian is lost. Still, it leaves the $\mathscr{H}^N_l$ subspaces uncoupled. In Fig.\,\ref{spectrumHS2HS4_eps0} we show the energy spectrum of Hamiltonian \eqref{fmham1_v2} for $N=2$ and $N=4$ in the symmetric subspace with $\epsilon=0$ and $U_2 = 0$, plotted against $\chi=\frac{2}{\pi}\arctan\left( \frac{U_1}{J} \right)$. The parameter $\chi$ ranges from the noninteracting scenario $\chi=0$ to the case with suppressed tunneling $\chi=1$. The spectrum exhibits avoided crossings with nonvanishing level repulsion across all the $\mathscr{H}^N$  subspaces. The level repulsion vanishes in all the crossings for $\epsilon=U_1/2$, as shown in Fig.\,\ref{spectrumHS2HS4_eps05}. The values of $\chi$ at which the crossings are found within a given subspace $\mathscr{H}^N$ are preserved along the subspaces $\mathscr{H}^{N'}$ with a higher number of particles $N'>N$ of the same number-parity, as illustrated for the two lowest eigenstates in Fig.\,\ref{spectrumHS2HS4_eps05}(c). This results from the block diagonalization of the Hamiltonian into the different angular momentum representations $\mathscr{H}_l^N$ at $\epsilon=U_1/2$. A similar behaviour is observed in the two-mode Bose Hubbard model with atom-pair tunneling along the boundary between phase-locking and self-trapping phases \cite{Rubeni-2017, Agboola-2018}. Indeed, rotating the basis \eqref{transfxy} by $\pi/2$ and setting $U_2=0$, we have
\small
\begin{align}\label{2xTMBH}
\ham_{t.b.} = \,&U_1\left(\hat{N}_x\hat{N}_M + \hat{N}_y\hat{N}_M \right) + \epsilon\hat{N}_M \nonumber\\
+&\frac{U_1}{2}\left( (\hat{b}_x^{\dagger})^2\hat{b}_M^2 + \hat{b}_x^2(\hat{b}_M^{\dagger})^2  + (\hat{b}_y^{\dagger})^2\hat{b}_M^2 + \hat{b}_y^2(\hat{b}_M^{\dagger})^2  \right) \nonumber\\
-&\frac{J}{2}\left(\hat{b}_x^{\dagger}\hat{b}_M + \hat{b}_x\hat{b}_M^{\dagger}  + \hat{b}_y^{\dagger}\hat{b}_M + \hat{b}_y\hat{b}_M^{\dagger}  \right).
\end{align}
\normalsize
Clearly, the tight-binding Hamiltonian \eqref{fmham1_v2} can be interpreted as the composition of two two-mode Bose Hubbard systems sharing one mode, where nonlinear atom-pair tunnelings are included. 

Unlike with the tunneling operator $\hat{J}$, the addition of a nonzero term $\ham_2$ \eqref{U2operator} breaks the $\text{U}(1)$ symmetry associated to the charge $l$ in the three mode Hamiltonian \eqref{fmham1_v2}. In the rotated basis:
\begin{equation}\label{U2hamrotated}
\ham_2 = \frac{U_2}{4}\left((\hat{N} - \hat{N}_M)^2 - \hat{L}_z^2\right).
\end{equation}

The term proportional to $\hat{N}_M$ supposes an effective decrease in the central well energy $\Delta\epsilon_M = -\frac{U_2}{2}N$, which introduces a shift proportional to the number of particles to the gap-closing condition for $\epsilon$. Yet the term $\frac{U_2}{4}\hat{N}_M^2$, being quadratic in the number operator $\hat{N}_M$, cannot be compensated by adjusting the single particle parameters, thus coupling the different subspaces $\mathscr{H}^N_l$ across all parameter space. However, its effect remains small in the regime with $\Omega < E_R$. Moreover, it is worth mentioning that while the total "angular momentum" $l$ is not preserved by $\ham_2$, the magnetization $m_z$ is.

\subsection{Dynamical properties of \texorpdfstring{$\ham_{t.b.}$}{}}\label{section4.2}

A trademark of the Hamiltonian derived in this work is the emergence of effective spin-changing collisions that couple the edge well states with the central one, described by the effective Hamiltonian $\ham_1$ \eqref{Uoperator}. These processes prevent the Fock basis states of the lowest band modes to be an eigenstate of the system, and can give rise to nontrivial dynamics when the interactions dominate over the noninteracting trapping-mediated tunneling dynamics. In the regime where the tunneling $\hat{J}$ is suppressed and the Raman coupling is weak, the dynamics is essentially described by $\ham_1$. As indicated in the previous section, such Hamiltonian is block-diagonalized in subspaces with preserved effective magnetization, $m_z$. Remarkably, in the subspace of zero magnetization we have
\begin{align}\label{fmham_mz0}
\ham_{1} \overset{m_z = 0} {\longrightarrow} \frac{U_1}{2}\left[\hat{L}^2 - \hat{N} + \left(\frac{2\epsilon}{U_1} -1\right)\hat{N}_M\right].
\end{align}
\begin{figure}[t]
\includegraphics[width=.80\linewidth]{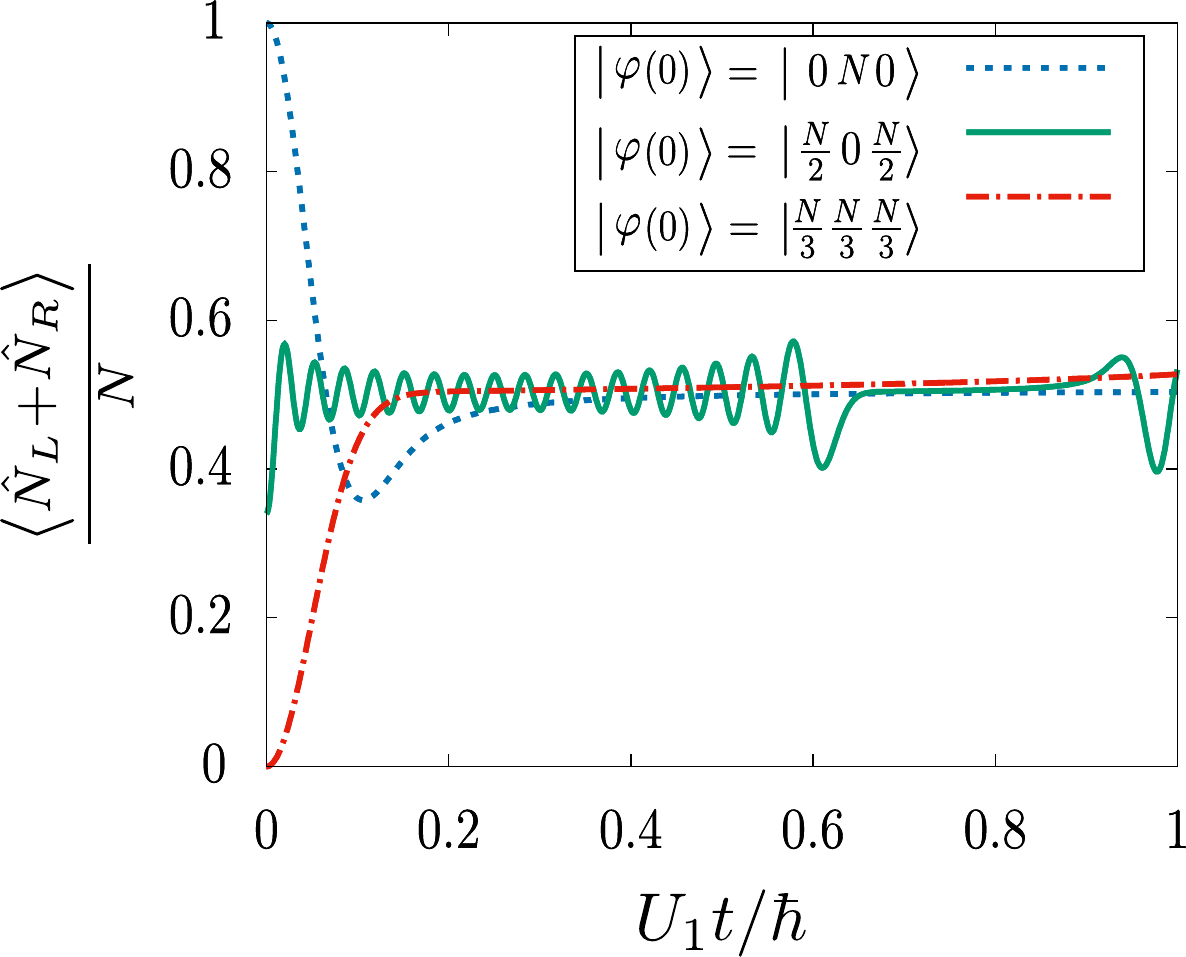}
\caption{\small (Color online) \textbf{Spin mixing induced by effective spin-dependent collisions:} Mean population in $\ket{M}$ as a function of time for a state initially prepared within the $m_z = 0$ subspace at $\ket{0N0}$ (dashed blue), $\ket{\frac{N}{3} \frac{N}{3} \frac{N}{3}}$  (solid green) and $\ket{\frac{N}{2} 0 \frac{N}{2}}$ (dashed-dotted red). The initial state is evolved under Hamiltonian \eqref{fmham1_v2} for $J = U_2 = 0$ and $\epsilon=U_1/2$. Time is scaled to $\hbar/U_1$.} 
\label{spinmixingdynam} 
\end{figure}

\begin{figure*}[t]
\includegraphics[width=.99\linewidth]{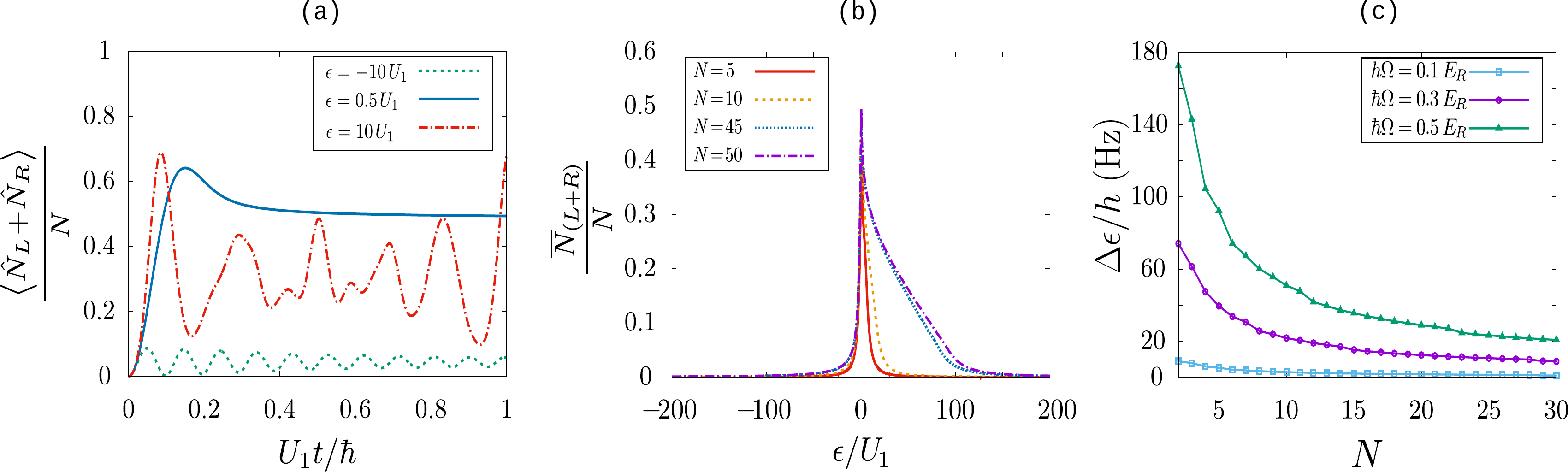}
\caption{\small (Color online) \textbf{Signature of the spin-mixing dynamics:} \textbf{(a)} Mean relative population in the edge wells, $\mean{\hat{N_L} + \hat{N_R}}$ as a function of time for a state initially prepared at $\ket{0 N 0}$, with $N= 50$, and evolved under Hamiltonian \eqref{fmham1_v2} for $J = U_2 = 0$ and $\epsilon=-10 U_1$ (dashed green), $\epsilon=0.5 U_1$ (solid blue) and $\epsilon=10 U_1$ (dashed-dotted red). Time is scaled to $\hbar/U_1$. \textbf{(b)} Time-averaged relative value of $\mean{\hat{N_L} + \hat{N_R}}$, as a function of $\epsilon$, averaged over large times, $\Delta t \sim 1/U_1$, for different numbers of particles: $N=5$ (solid red), $N=10$ (dashed orange), $N=45$ (dotted blue) and $N=50$ (dashed-dotted purple). \textbf{(c)} Width $\Delta\epsilon$ of the peak as a function of the total number of particles $N$ for $\alpha = 0.1$ with optimal parameters and considering $^{87}$Rb atoms. $\hbar\Omega/E_R$ is set to $0.5$, $0.3$ and $0.1$, plotted in squared blue, circular red and triangular green dots, respectively.}\label{quencheddynam} 
\end{figure*}

which is analogous to the Hamiltonian describing the spin dynamics of a spinor BEC with spin-dependent collisions \cite{Law-1998, Zhang-2005,Chang-2005}. Like in such system, here an arbitrary state initially prepared in the $m_z = 0$ manifold undergoes nonlinear coherent spin mixing when evolved under $\ham_1$, as illustrated in Fig.\,\ref{spinmixingdynam}, over times characterized by
\begin{equation}
\tau_c \approx \frac{\hbar}{\sqrt{N} U_1}.
\end{equation}

However, here the coherent mixing is strongly dependent on the resonant condition at $\epsilon=U_1/2$, with the amplitude of the spin oscillations decreasing as $\epsilon$ departs from $U_1/2$. This is illustrated in Fig.\,\ref{quencheddynam}.(a) and Fig.\,\ref{quencheddynam}.(b) for a state initially prepared at $\ket{N_L N_M N_R} = \ket{0N0}$. The mean relative population in the edge wells, $\mean{\hat{N_L} + \hat{N_R}}$ is plotted as a function of time in Fig.\,\ref{quencheddynam}.(a) for different values of $\epsilon$. Its time-averaged value, which we label as  $\overline{N}_{\small{(L+R)}}$, is represented in Fig.\,\ref{quencheddynam}.(b) for different number of particles, averaged over long times ($\Delta t \sim \hbar/U_1$). The shape of the peak converges fast as $N$ is increased. As expected, the maximum converges to $1/2$ at $\epsilon=U_1/2$, when the spin-mixing is the largest.   

In an experimental implementation, the state $\ket{0N0}$ can be easily prepared by initially setting $\epsilon \ll -1$, followed by a quench in the central well energy $\Delta_0$ (as defined in section \ref{section2.1}) to reach the targeted value of $\epsilon$. The quench can be performed without considerably populating the higher bands due to the scale separation between the gap $\propto \Omega$ and $U_1\ll\hbar \Omega$. 
With this preparation, the time-averaged value $\overline{N}_{\small(L+R)}$ is a suitable observable to probe experimentally the correlated spin dynamics induced by the tight-binding Hamiltonian \eqref{fmham1_v2}. The peak that can be observed in the edge-wells population around the resonant condition (as shown in Fig.\,\ref{quencheddynam}.(b)) is robust due to the dynamics being insensitive to the fluctuations of the magnetic field at linear Zeeman level. This is clear as we prepare the initial state in the zero magnetization manifold. Furthermore, a large energy separation between the subspaces with different magnetization $m_z$ can be induced by having a relatively large detuning $\delta = \abs{\Delta_{L} - \Delta_{R}}$, which downplays the noncoherent population of the edge wells during the time evolution. Still, we need to account for the quadratic contribution to the Zeeman split, albeit it is typically much smaller.

To assess the visibility of the spin dynamics we compare the experimental error associated to $\Delta_0$ to a realistic estimation of the width of the resonance peak around $\epsilon=U_1/2$, that we characterize by the variance $\Delta\epsilon$, taking $\overline{N}_{\small(L+R)}(\epsilon)$ as a distribution. As shown in Fig.\,\ref{quencheddynam}(b), $\Delta\epsilon/U_1$ converges fast when $N$ is increased. The width of the peak is proportional to the interacting coefficient $U_1$, which, from equations \eqref{Ucoefficients},\eqref{ccoeff},\eqref{U0coeff} and \eqref{Uijk}, reads
\begin{equation}
U_1 \simeq  \frac{g C_1\mean{n}}{N} \label{omega1density},
\end{equation}
where $C_1 := C_{1(-1)00}(\Omega)$ (recall \eqref{ccoeff}), with $0\leq C_1 \leq 1$. The function $C_1$ depends strongly on $\Omega$, but it does not vary significantly with respect to $\omega_z$ and $\Delta_i$ in the regimes we consider. Recall from \eqref{densityupbound2} that the validity of the tight-binding Hamiltonian \eqref{fmham1_v2} established an upper bound on the atom density, and thus on the interaction coefficient $U_1$. By parametrizing
\begin{align}
g\mean{n} &= \alpha\hbar\omega_z, \text{\quad with }0 < \alpha \ll 1,\\
\hbar\omega_z &= \beta E_R, \text{\quad with }0 < \beta \ll 1,
\end{align}
it follows from \eqref{omega1density}
\begin{equation}\label{U1ofEr}
U_1 \simeq  \frac{\alpha\beta C_1}{N} E_R \ll \frac{C_1}{N} E_R.
\end{equation}

In this way, from \eqref{U1ofEr} it is clear that $\Delta\epsilon$ is a decreasing function of $N$. Notice how its expression depends only on the properties of the Raman couplings, from which the constraints $\hbar\omega_z \ll E_R$ and $\hbar\Omega < E_R$ stem. The optimal value for $\Delta\epsilon$ is retrieved by optimizing the factor $\alpha\beta C_1$, given the constraints assumed in the derivation of the effective Hamiltonian \eqref{fmham1_v2}
\begin{align}
\abs{U_{(j+k)(i-l)ij}} &\ll \abs{U_{mn}^{(1)}}, \text{ for } k \neq l, \text{ all }i,j,m,n,\label{deltazconst2} \\
\abs{U_{ij}^{(0)} - U_{kl}^{(0)}} &\ll \abs{U_{mn}^{(1)}}, \text{ for all } i,j,k,l,m,n. \label{deltazconst3}
\end{align}
Furthermore, in order to confine the many-body state within the $m_z = 0$ subspace during the dynamics, we require that
\begin{equation}\label{deltazconst1}
\abs{J} \ll \abs{U_1}.
\end{equation}
For given $\Omega$ and $\beta$, $\alpha$ can be independently tuned by adjusting the transverse confinement, i.e. $\hat{V}_{\perp}$, while making sure that $\alpha \ll 1$ to be well within the tight-binding approximation. In Fig.\,\ref{quencheddynam}(c), we plot the width of the resonance peak, $\Delta\epsilon$, as a function of the total number of particles $N$, for $\alpha=0.1$ and for different values of $\Omega$. There, the value of $\beta$ is numerically optimized at each $\Omega$ and $N$, given the discussed constraints: the terms that appear on the left hand side of \eqref{deltazconst2}, \eqref{deltazconst3} and \eqref{deltazconst1} are constrained to be smaller than $10^{-2}\abs{U_1}$ in the numerical calculations. We have set $E_R$ to its typical value in $^{87}\text{Rb}$ SOC experiments at around $E_R/\hbar \sim 2\pi\cdot 1.5\cdot10^4$ $\text{Hz}$. For comparison, we now consider the experimental values employed by Campbell \textit{ et al.} in a spin-1 SOC Bose gas experiment with $^{87}\text{Rb}$ \cite{Campbell-2016}. There, in order to tune $\Delta_0>0$ in Hamiltonian \eqref{hamspnotrap2}, each pair of hyperfine states is coupled by independent Raman transitions. This requires that the quadratic Zeeman split is large enough so that each Raman pair only resonantly couples one transition. There, such effect is achieved with static bias fields in the order of few tens of Gauss, for which the resulting linear Zeeman split is still much more significant. With an error in the quadratic Zeeman splitting of just $1$ Hz, the main contribution to $\delta_{\text{R}}$ in the experiment stems from the residual cross coupling between both $\ket{1,-1}$ to $\ket{1,0}$, and $\ket{1,0}$ to $\ket{1,1}$ transitions. The effect of such cross coupling, calculated from Floquet theory, depends on the Rabi frequency of the transitions.  We calculate its associated error from \cite[eq. 8]{Campbell-2016}, considering $\hbar\Omega = 0.50(1) E_R$, to be around $5$ Hz, way below the corresponding values of $\Delta\epsilon$ that can be obtained in the few particle regime for moderate values of $\Omega$ (see Fig.\,\ref{quencheddynam}(c)). Therefore, the spin-mixing dynamics is resolvable in realistic experiments. We note that the different sources of noise could in principle be optimized further. For instance, recently, in \cite{Xu-2019}, the root mean square value of the magnetic field noise was kept as low as few tens of $\mu$G.

Finally, we need to consider that Raman driven gases are subjected to strong heating, which severely limit the coherence lifetime. With the preparation suggested in this section, where all the atoms are initially loaded in the $\ket{0N0}$ state, the characteristic time of the relaxation process scales with $N^{1/2}$:
\begin{equation}
\tau_c \approx \sqrt{N}\frac{\hbar}{U_1N} = \sqrt{N}\frac{\hbar}{\alpha \beta C_1 E_R} \sim \sqrt{N}10^{-2} \text{ s.}      
\end{equation}
In the few particle regime the coherent evolution time required are of the order of $10$ ms. Interestingly, as the initial spin mixture is increased, the frequency of the coherent oscillations $f_c$ increases, reaching $f_c \sim \frac{\alpha \beta C_1 E_R}{\hbar} \sim 10^2$ Hz when the initial state is maximally mixed, as we show in Fig.\,\ref{spinmixingdynam} for $N=100$. Naturally, the amplitude of the oscillations diminishes, yet this allows the initial mixing to be optimized when constrained by the coherence lifetime in the gas. As an estimation of the heating in the gas, we consider the lifetime of the spin-orbit coupled BEC, which can currently be extended up to $1$s \cite{Luo-2016, Wu-2016, Sun-2016}. In the scheme proposed, starting from a BEC prepared in the $F=1$ manifold, the targeted Rabi frequency and Raman detunings to prepare the pre-quench initial state can be adiabatically achieved in less than $100$ ms \cite{Williams-2012, Sun-2016}. This is followed by the quench in $\epsilon$, performed in comparably negligible time below the ms. After the quench, the system is left to evolve for a time interval of the order of $\tau_c$. A second quench back to the pre-quench conditions, at $-\epsilon/U_1 \gg 1$, can be then applied to freeze the interaction-driven dynamics. Finally, the Rabi frequency can be adiabatically turned off in order to gain a strong correlation between the occupation of well states and the corresponding spin states, which is partly lost at larger $\Omega$, achieving better resolution in the eventual Stern-Gerlach measurement of the populations. Altogether, the protocol could be performed in less than $300$ ms. With these prospects, while challenging, we find the measurement feasible in the regimes suggested. As a final remark, we note that the few particle regime can be explored by loading a very dilute gas into a two-dimensional lattice so as to have a low number of atoms per site.

\section{Conclusion}

In this paper, we have explored beyond-mean-field properties of spin-1 Bose gases with Raman driven SOC at low energies. The spin-texture in the single-particle dispersion bands that emerges from SOC modulates the amplitudes of the scattering processes in the gas. Following a tight-binding approximation at weak Raman coupling, where the lowest band presents a triple-well shape, we have shown that such modulation leads to effective correlated tunneling processes between the site modes in momentum space. Their presence supposes a departure from conventional position space analogies. We have discussed the spectral properties of the Hamiltonian, showing that it becomes integrable in a certain region of the parameter space. In such conditions, the gas undergoes interaction-driven coherent spin dynamics, similarly to what occurs in spinor BECs with spin-dependent scattering parameters \cite{Law-1998, Zhang-2005,Chang-2005}, with spin mixing occurring over sufficiently large times. Remarkably, we have shown that such beyond-mean-field effects can dominate the dynamics of the system in the few-particle regime. Exploiting the  difference between the noninteracting and the interacting energy scales, we have proposed a quench protocol through which all the noninteracting dynamics is frozen. The visibility of the induced spin dynamics heavily relies on the choice of the experimental signature and the initial state, which allows the protocol to be robust against relatively large fluctuations in the bias magnetic field. Finally, we have shown how the predicted spin dynamics can be measured in state-of-the-art experiments.

The spin-1 case we have discussed here is the minimal spin size where the described beyond-mean-field effects manifest and can be experimentally detected as spin-changing collisions. Similar and more complex processes appear at higher spins. Such terms can take place in spin-orbit coupled Bose gases of alkali atoms such as Caesium ($F=3$) or Lanthanide atoms such as Dysprosium \cite{Burdick-2016}. Furthermore, we have shown the analogy of the system described to the spinor dynamics in spin-dependent interacting gases. This analogy suggests that synthetic spin-orbit coupling could be employed for the generation of  macroscopically entangled states, as in \cite{Zhang-2013} \cite{Luo-2017}, to be used in metrological applications. These aspects will be covered in an upcoming work \cite{Cabedo-2019-workinprogress}.

\acknowledgements
\small
J.Cabedo, A.C., V.A. and J.M. acknowledge supported from the Ministerio de Economía y Competividad MINECO (Contract No. FIS2017-86530-P) and from Generalitat de Catalunya (Contract No. SGR2017-1646). J.Claramunt was supported by DGI-MINECO-FEDER (Grants MTM2017-83487-P and BES-2015-071439). A.C. acknowledges support from the UAB Talent Research program. Y.Z. was supported by the NNSF of China (Grand No.11774219). The authors also thank M.Nakahara for insightful comments and useful discussions. 
\normalsize


%

\end{document}